\documentclass[aps,prl,reprint,showpacs,superscriptaddress]{revtex4-1}
\usepackage{amsmath}
\usepackage{SIunits}
\usepackage{graphicx}% Include figure files

\begin{document}
%
%
% Define new commands:
%
\newcommand{\ac}[0]{\ensuremath{\hat{a}_{\mathrm{c}}}}
\newcommand{\adagc}[0]{\ensuremath{\hat{a}^{\dagger}_{\mathrm{c}}}}
\newcommand{\aR}[0]{\ensuremath{\hat{a}_{\mathrm{R}}}}
\newcommand{\aT}[0]{\ensuremath{\hat{a}_{\mathrm{T}}}}
\renewcommand{\b}[0]{\ensuremath{\hat{b}}}
\newcommand{\bdag}[0]{\ensuremath{\hat{b}^{\dagger}}}
\newcommand{\betaI}[0]{\ensuremath{\beta_\mathrm{I}}}
\newcommand{\betaR}[0]{\ensuremath{\beta_\mathrm{R}}}
\renewcommand{\c}[0]{\ensuremath{\hat{c}}}
\newcommand{\cdag}[0]{\ensuremath{\hat{c}^{\dagger}}}
\newcommand{\CorrMat}[0]{\ensuremath{\boldsymbol\gamma}}
\newcommand{\Deltacs}[0]{\ensuremath{\Delta_{\mathrm{cs}}}}
\newcommand{\Deltacsmax}[0]{\ensuremath{\Delta_{\mathrm{cs}}^{\mathrm{max}}}}
\newcommand{\Deltacsparked}[0]{\ensuremath{\Delta_{\mathrm{cs}}^{\mathrm{p}}}}
\newcommand{\Deltacstarget}[0]{\ensuremath{\Delta_{\mathrm{cs}}^{\mathrm{t}}}}
\newcommand{\Deltae}[0]{\ensuremath{\Delta_{\mathrm{e}}}}
\newcommand{\Deltahfs}[0]{\ensuremath{\Delta_{\mathrm{hfs}}}}
\newcommand{\dens}[0]{\ensuremath{\hat{\rho}}}
\newcommand{\erfc}[0]{\ensuremath{\mathrm{erfc}}}
\newcommand{\Fq}[0]{\ensuremath{F_{\mathrm{q}}}}
\newcommand{\gammapar}[0]{\ensuremath{\gamma_{\parallel}}}
\newcommand{\gammaperp}[0]{\ensuremath{\gamma_{\perp}}}
\newcommand{\gavg}[0]{\ensuremath{\mathcal{G}_{\mathrm{avg}}}}
\newcommand{\gbar}[0]{\ensuremath{\bar{g}}}
\newcommand{\gens}[0]{\ensuremath{g_{\mathrm{ens}}}}
\renewcommand{\H}[0]{\ensuremath{\hat{H}}}
\renewcommand{\hbar}[0]{}
\renewcommand{\Im}[0]{\ensuremath{\mathrm{Im}}}
\newcommand{\kappac}[0]{\ensuremath{\kappa_{\mathrm{c}}}}
\newcommand{\kappamin}[0]{\ensuremath{\kappa_{\mathrm{min}}}}
\newcommand{\kappamax}[0]{\ensuremath{\kappa_{\mathrm{max}}}}
\newcommand{\ket}[1]{\ensuremath{|#1\rangle}}
\newcommand{\mat}[1]{\ensuremath{\mathbf{#1}}}
\newcommand{\mean}[1]{\ensuremath{\langle#1\rangle}}
\newcommand{\omegac}[0]{\ensuremath{\omega_{\mathrm{c}}}}
\newcommand{\omegas}[0]{\ensuremath{\omega_{\mathrm{s}}}}
\newcommand{\pauli}[0]{\ensuremath{\hat{\sigma}}}
\newcommand{\pexc}[0]{\ensuremath{p_{\mathrm{exc}}}}
\newcommand{\pexceff}[0]{\ensuremath{p_{\mathrm{exc}}^{\mathrm{eff}}}}
\newcommand{\Pa}[0]{\ensuremath{\hat{P}_{\mathrm{c}}}}
\newcommand{\Qmin}[0]{\ensuremath{Q_{\mathrm{min}}}}
\newcommand{\Qmax}[0]{\ensuremath{Q_{\mathrm{max}}}}
\renewcommand{\Re}[0]{\ensuremath{\mathrm{Re}}}
\renewcommand{\S}[0]{\ensuremath{\hat{S}}}
\newcommand{\Sminuseff}[0]{\ensuremath{\hat{S}_-^{\mathrm{eff}}}}
\newcommand{\Sxeff}[0]{\ensuremath{\hat{S}_x^{\mathrm{eff}}}}
\newcommand{\Syeff}[0]{\ensuremath{\hat{S}_y^{\mathrm{eff}}}}
\newcommand{\tildeac}[0]{\ensuremath{\tilde{a}_{\mathrm{c}}}}
\newcommand{\tildepauli}[0]{\ensuremath{\tilde{\sigma}}}
\newcommand{\Tcaveff}[0]{\ensuremath{T_{\mathrm{cav}}^{\mathrm{eff}}}}
\newcommand{\Techo}[0]{\ensuremath{T_{\mathrm{echo}}}}
\newcommand{\Tmem}[0]{\ensuremath{T_{\mathrm{mem}}}}
\newcommand{\Tswap}[0]{\ensuremath{T_{\mathrm{swap}}}}
\newcommand{\Var}[0]{\ensuremath{\mathrm{Var}}}
\renewcommand{\vec}[1]{\ensuremath{\mathbf{#1}}}
\newcommand{\Xa}[0]{\ensuremath{\hat{X}_{\mathrm{c}}}}

\title{Quantum memory for microwave photons in an inhomogeneously
  broadened spin ensemble}

\author{Brian Julsgaard}
\email{brianj@phys.au.dk}
\affiliation{Department of Physics and Astronomy, Aarhus University, Ny
  Munkegade 120, DK-8000 Aarhus C, Denmark.}

\author{C\'ecile Grezes}
\author{Patrice Bertet}
\affiliation{Quantronics group, SPEC (CNRS URA 2464), IRAMIS, DSM,
  CEA-Saclay, 91191 Gif-sur-Yvette, France}

\author{Klaus M{\o}lmer}
\affiliation{Department of Physics and Astronomy, Aarhus University, Ny
  Munkegade 120, DK-8000 Aarhus C, Denmark.}


\date{\today}

\begin{abstract}
  We propose a multi-mode quantum memory protocol able to store the
  quantum state of the field in a microwave resonator into an ensemble
  of electronic spins. The stored information is protected against
  inhomogeneous broadening of the spin ensemble by spin-echo
  techniques resulting in memory times orders of magnitude longer than
  previously achieved. By calculating the evolution of the first and
  second moments of the spin-cavity system variables for realistic
  experimental parameters, we show that a memory based on NV center
  spins in diamond can store a qubit encoded on the $\ket{0}$ and
  $\ket{1}$ Fock states of the field with 80\% fidelity.
\end{abstract}


\maketitle

Ensembles of electronic spins have been proposed as quantum memories
in hybrid architectures for quantum computing including
superconducting qubits \cite{Imamoglu.PhysRevLett.102.083602(2009),
  Wesenberg.PhysRevLett.103.070502(2009),
  Marcos.PhysRevLett.105.210501(2010),
  Yang.PhysRevA.84.010301R(2011)}. Progress in this direction was
reported in a number of experiments, demonstrating first strong
coupling of an ensemble of spins in a crystal to a superconducting
resonator \cite{Kubo.PhysRevLett.105.140502(2010),
  Schuster.PhysRevLett.105.140501(2010),
  Amsuss.PhysRevLett.107.060502(2011),
  Staudt.JPhysBAtMolOptPhys.45.124019(2012),
  Huebl.arXiv.1207.6039(2012), Ranjan.arXiv.1208.5473(2012),
  Probst.arXiv.1212.2856(2012)}, and more recently reversible storage
of a single microwave photon in the spin ensemble
\cite{Kubo.PhysRevLett.107.220501(2011),
  Zhu.Nature.478.221(2011)}. From these results it clearly appears
that inhomogeneous broadening of the spin ensemble is a major
obstacle, which needs to be overcome for hybrid quantum circuits to
fully benefit from the long spin-coherence times. Due to inhomogeneous
broadening, quantum information leaks from the "bright" collective
degree of freedom coupled to the cavity into dark modes of the spin
ensemble \cite{Kurucz.PhysRevA.83.053852(2011),
  Diniz.PhysRevA.84.063810(2011), Kubo.PhysRevA.85.012333(2012)}. An
appealing possibility is to actively and coherently restore it using
refocusing techniques, inspired from magnetic-resonance methods
\cite{Hahn.PhysRev.80.580(1950)} and based on the application of $\pi$
pulses to the spins acting as time reversal. However, these ideas face
a number of challenges: (i) The spatial inhomogeneity of the microwave
resonator field may make it difficult to apply a $\pi$ pulse
efficiently to each spin, (ii) after the $\pi$-pulse inversion, the
spin ensemble should remain stable despite its coupling to the cavity,
and (iii) the whole statistics of the collective spin must be restored
at the single quantum level. The present work proposes a protocol,
which addresses all these issues, and we exemplify its feasibility for
the specific case of NV centers in diamond
\cite{Gruber.Science.276.2012(1997)}, using currently available
experimental techniques and realistic parameters. The proposed memory
extends the storage times by several orders of magnitude compared to
\cite{Kubo.PhysRevLett.107.220501(2011), Zhu.Nature.478.221(2011)}. It
is intrinsically multi-mode and thus allows to store reversibly a
number of quantum states, paving the way to the realization of a
genuine quantum Turing machine
\cite{Tordrup.PhysRevLett.101.040501(2008),
  Wesenberg.PhysRevLett.103.070502(2009)}.

In our proposal the $\pi$ pulses are performed by rapid adiabatic
passage \cite{Abragam.PrinciplesNucMag(1961)} through the electron
spin resonance, a method known to tolerate an inhomogeneous microwave
field. Stability of the ensemble after inversion is ensured provided
the cavity quality factor is sufficiently low
\cite{Julsgaard.PhysRevA.86.063810(2012)}. Since this is incompatible
with a faithful transfer of quantum information from the cavity into
the spins, we propose to use a cavity with a quality factor that can
be tuned in-between the steps of the protocol, as was recently
demonstrated with SQUIDs
\cite{Palacios-Laloy.JLowTempPhys.151.1034(2008)}. In addition,
inspired by a recent proposal of atomic-ensemble quantum memories for
optical photons \cite{Damon.NewJPhys.13.093031(2011)}, we employ two
$\pi$ pulses in the refocusing scheme. To avoid emitting a microwave
echo from the inverted spin ensemble, which would otherwise be more
noisy than the original quantum state
\cite{Ruggiero.PhysRevA.79.053851(2009)}, we detune the cavity from
the spins in-between the two pulses (effectively "silencing" this
noisy first echo \cite{Damon.NewJPhys.13.093031(2011)}). The second
echo, formed in a non-inverted ensemble, restores quantum information
with a fidelity up to 80 \% for realistic parameters.
%We develop novel numerical methods to calculate the evolution of the
%quantum state of spins and cavity under the influence of
%inhomogeneities and arbitrary spin excitation allowing us to show that
%the second echo reproduces the quantum statistics of the initial
%field.
%
%
\begin{figure}[t]
  \centering
  \includegraphics{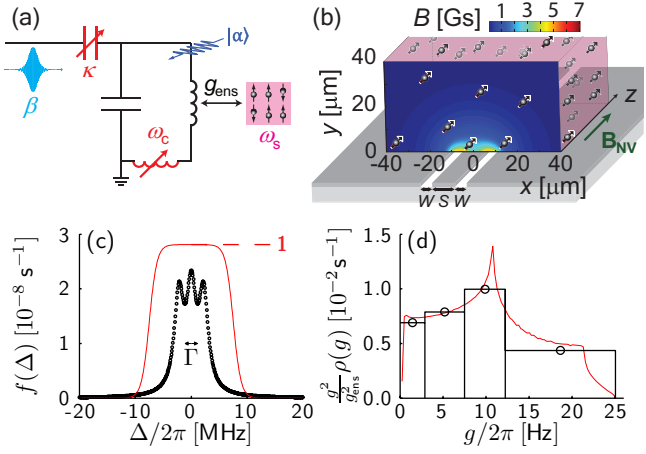}
  \caption{(Color online) (a) Quantum memory circuit. The resonator,
    with frequency $\omegac$ and damping rate $\kappa$ tunable at the
    nanosecond scale, is coupled to the spin ensemble (frequency
    $\omegas$) with an ensemble coupling constant $\gens$. Drive
    pulses of amplitude $\beta(t)$ are applied to the spins via the
    resonator, which can be initialized in a well-defined quantum
    state $\ket{\alpha}$. (b) Amplitude of the microwave field
    generated by a coplanar resonator with quality factor $Q=100$ and
    driven by a pulse of 100 {\micro}W power. A static magnetic field
    $\vec{B}_{\mathrm{NV}}$ is applied parallel to the spins, which
    are distributed uniformly throughout the crystal. (c) Sub-ensemble
    distribution $f(\Delta)$ of spin-resonance frequencies (circles)
    consisting of three hyperfine-split Lorentzian lines. The solid
    line shows the excitation probability for the chosen secant
    hyperbolic inversion pulses (see text). (d) (Solid line)
    Calculated coupling-strength distribution function
    $\rho(g)g^2$. (Histogram and circles) Sub-ensemble distribution
    used in the calculation. The low- and high-frequency cut-offs in
    $\rho(g)$ originate from, respectively, high
    ($40\:\micro\mathrm{m}$) and low ($0.5\:\micro\mathrm{m}$)
    cut-offs in the distance from the resonator to NV centers.}
\label{fig:Setup}
\end{figure}

The proposed physical setup is shown schematically in
Figs.~\ref{fig:Setup}(a) and (b); a diamond crystal containing NV
centers \cite{Gruber.Science.276.2012(1997)} is placed on top of a
transmission-wave-guide cavity whose frequency $\omegac$
\cite{Palacios-Laloy.JLowTempPhys.151.1034(2008)} and coupling to the
measuring line $\kappa$ \cite{Yin.ArXiv.1208.2950(2012)} can be tuned
on a nanosecond timescale using control lines (not shown in
Fig.~\ref{fig:Setup}). The crystal is subjected to a constant bias
magnetic field $\vec{B}_{\mathrm{NV}}$, lifting the degeneracy of the
$m_{\mathcal{S}} = \pm 1$ states, and bringing the $0 \rightarrow 1$
transition to an average frequency $\omegas= 2\pi\cdot
2.9\:\mathrm{GHz}$. In the frame rotating at $\omegas$ the free
evolution of the cavity field and the spin ensemble is described by
the Hamiltonian: $\H_0 = \hbar\Deltacs\adagc\ac +
\hbar\sum_j\frac{\Delta_j}{2}\pauli_z^{(j)}$, where $\ac$ is the
cavity field annihilation operator, $\Deltacs = \omegac - \omegas$ is
the (adjustable) spin-cavity detuning, $\Delta_j = \omega_j -
\omegas$, $\omega_j$ the resonance frequency of the $j$th spin and
$\pauli_z^{(j)}$ the corresponding Pauli operator. NV centers are
coupled by hyperfine interaction to the nuclear spin of their nitrogen
atom (having a spin 1), causing the $m_{\mathcal{S}} = 0\rightarrow 1$
transition to split into a triplet separated by $\Deltahfs/2\pi =
2.2\:\mathrm{MHz}$ \cite{Felton.PhysRevB.79.075203(2009)}. In
addition, they are coupled by dipolar interactions to a bath of
magnetic dipoles \cite{SuppMater}, which is known to govern their
coherence time \cite{Hanson.Science.320.352(2008),
  Dobrovitski.PhysRevB.77.245212(2008),
  Zhao.PhysRevB.85.115303(2012)}. This bath broadens each of the
hyperfine resonances, with a Lorentzian line shape
\cite{Dobrovitski.PhysRevB.77.245212(2008)} of width $w$,
corresponding to a free-induction-decay time $T_2^*= \frac{2}{w}$. A
Hahn-echo pulse sequence \cite{Hahn.PhysRev.80.580(1950)} partially
refocuses this coherence, yielding a coherence time $T_2$ which can be
several orders of magnitude longer than $T_2^*$. In this work, we thus
model the system by the static inhomogeneous spin distribution shown
in Fig.~\ref{fig:Setup}(c) of characteristic width $\Gamma \approx w$
\cite{SuppMater}, and damped at a rate $\gammaperp = T_2^{-1}$ in the
Markov approximation \cite{SuppMater}. The spin-cavity interaction is
described by: $\H_{\mathrm{I}} = \hbar\sum_j
g_j(\pauli_+^{(j)}\ac+\pauli_-^{(j)}\adagc)$, where the coupling
constant $g_j$ of the $j$th spin is distributed as shown in
Fig.~\ref{fig:Setup}(d) \cite{SuppMater}. This distribution is of no
concern for storing the quantum state
\cite{Kubo.PhysRevLett.107.220501(2011)}; however, it prevents the
application of a "hard" $\pi$ pulse since each spin has a different
Rabi frequency for a given drive amplitude. So-called hyperbolic
secant pulses \cite{Silver.PhysRevA.31.R2753(1985)}, where the pulse
amplitude and phase are modulated as $a_{\mathrm{c}} =
a_{\mathrm{c}}^{\mathrm{max}} [\mathrm{sech}
(\beta_{\mathrm{sech}}t)]^{1+i\mu}$, are known to remedy this issue
\cite{Garwood.JMagRes.153.155(2001)}.  The pulses are applied by an
external drive $\beta$ modeled by the Hamiltonian $\H_{\mathrm{ext}} =
i\hbar\sqrt{2\kappa}(\beta \adagc - \beta^*\ac)$. Note that to achieve
the desired temporal dependence of $\ac$, $\beta$ must be further
tailored in order to account for the cavity filtering and the coupling
to the spins \cite{SuppMater}.

\begin{figure}[t]
  \centering
  \includegraphics{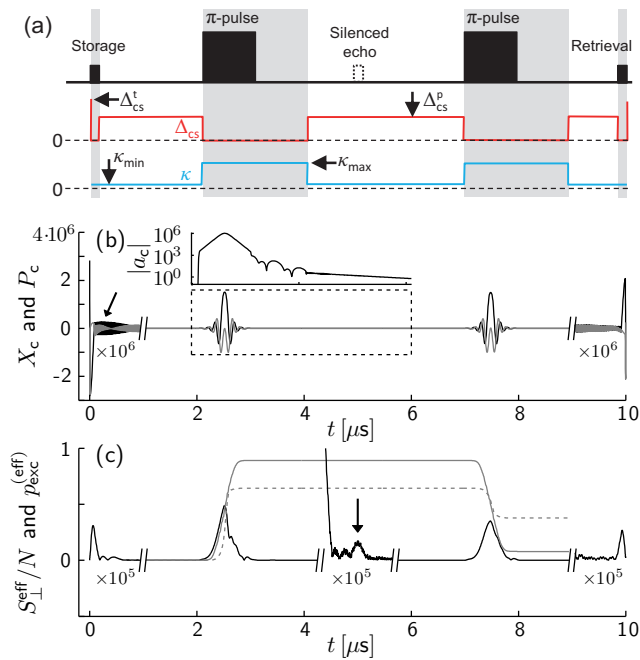}
  \caption{(Color online) (a) Schematic timing of pulses and cavity
    parameters, $\Deltacs$ and $\kappa$. Periods of resonance
    ($\Deltacs = 0$) are marked by gray areas. (b) Cavity-field mean
    values, $X_{\mathrm{c}}$ (black) and $P_{\mathrm{c}}$ (gray)
    versus time. The inset re-plots the dashed-line region with
    $|\mean{\ac}|$ on the logarithmic vertical scale. (c) The
    $g$-weighted transverse-spin-component mean value
    $S_{\perp}^{\mathrm{eff}} = \sqrt{S_x^{\mathrm{eff\,2}} +
      S_y^{\mathrm{eff\,2}}}$ (black) normalized to $N$, the
    excitation probability $\pexc$ (gray, dashed curve), and the
    $g$-weighted excitation probability $\pexceff$ (gray, solid
    curve).}
\label{fig:MeanValues}
\end{figure}

The quantum memory protocol, shown schematically in
Fig.~\ref{fig:MeanValues}(a), aims to store a cavity-field state given
at $t=0$ and retrieve it again at $t = \Tmem$ with the cavity tuned to
a ``target frequency'' $\Deltacstarget$. This quantum state could be
delivered by, e.g., a super-conducting transmon qubit along the lines
of \cite{Kubo.PhysRevLett.107.220501(2011)}. The cavity state is then
transfered to the spins by setting $\Deltacs = 0$ for a time $\Tswap$
after which the cavity is parked at $\Deltacs^{\mathrm{p}}$.  In a
lowest-order approximation $\Tswap = \pi/2\gens$ where $\gens = [\int
g^2\rho(g)dg]^{1/2}$ corresponds to the resonator-spin ensemble swap
rate \cite{Kubo.PhysRevLett.107.220501(2011),
  Zhu.Nature.478.221(2011), Kubo.PhysRevA.85.012333(2012)}, but in
reality is optimized numerically. For a high-fidelity storage, we set
$\kappa = \kappamin = \frac{\omegac}{2\Qmax}$ with $\Qmax = 10^4$ so
that the spin ensemble and resonator are in strong coupling. Next, in
order to refocus the reversible spin dephasing we apply two $\pi$
pulses at $\sim\!\frac{\Tmem}{4}$ and $\sim\!\frac{3\Tmem}{4}$ with
$\Deltacs = 0$, and to stabilize the inverted spin ensemble, we set
$\kappa = \kappamax = \frac{\omegac}{2\Qmin}$ with $\Qmin = 100$
before the $\pi$ pulses so that the cooperativity parameter fulfills
$C = \frac{\gens^2}{\kappa\Gamma} < 1$
\cite{Julsgaard.PhysRevA.86.063810(2012)}.  An additional constraint
comes from the fact that tuning the cavity frequency or quality factor
with SQUIDs is possible only if the cavity field is sufficiently low
($|\mean{\ac}| \lesssim 10$), which requires sufficient delay to allow
it to decay after the $\pi$ pulses. Between the two $\pi$ pulses, we
set $\Deltacs = \Deltacs^{\mathrm{p}}$ in order to silence the first
spin echo \cite{Damon.NewJPhys.13.093031(2011)}. After the second
$\pi$ pulse the quantum state is retrieved from the spin ensemble by
setting $\Deltacs = 0$ during $\Tswap$ after which the cavity is tuned
to $\Deltacstarget$.

The numerical calculation of the dynamical evolution is made tractable
by dividing the spins into $M$ sub-ensembles along the lines of
\cite{Julsgaard.PhysRevA.86.063810(2012)} keeping account of the mean
values and covariances between cavity-field quadratures, $\Xa$ and
$\Pa$, and spin components, $\S_x^{(m)}$, $\S_y^{(m)}$, and
$\S_z^{(m)}$ of the $m$th sub-ensemble \cite{SuppMater}. Such a
representation is convenient for determining the memory performance
for, e.g., coherent input states. Specific for our NV-center example
we use $\gens = 2\pi\cdot 3.5\:\mathrm{MHz}$, $w = 2\pi\cdot
2\:\mathrm{MHz}$ corresponding to $T_2^* = 0.16\:\micro\mathrm{s}$,
$T_2 = 100\:\micro\mathrm{s}$ \cite{SuppMater}, and hyperbolic secant
$\pi$ pulses truncated at a duration of 1 {\micro}s with $\mu = 3.5$
and $\mu\beta_{\mathrm{sech}} = 2\pi\cdot 7.5\:\mathrm{MHz}$. We
assume that a microwave drive of peak power up to 100 {\micro}W can be
applied to the sample input without causing too much heating.

Typical results of our calculations are shown in
Fig.~\ref{fig:MeanValues}. Panel (b) shows the mean values of $\Xa$
and $\Pa$ when a weak coherent cavity-field state is given at
$t=0$. Even though the cavity field is very strong during the
inversion pulses at $t\approx 2.5\:\micro$s and $t\approx
7.5\:\micro$s, it relaxes to negligible levels prior to memory
retrieval. Due to an imperfect storage process [marked by the arrow in
panel (b)] a minor part of the field is left in the cavity (14 \% in
field strength or 2 \% in energy units), but most importantly
$|\mean{\ac}|$ recovers at $t=\Tmem$ a value comparable to the one at
$t=0$. Regarding the spin state, we consider the \emph{effective},
$g$-weighted spin observables, $\S_{\eta}^{\mathrm{eff}} = \sum_j
\frac{g_j}{\gbar}\pauli_{\eta}^{(j)}$, $\eta=x,y$ and $\gbar =
\gens/\sqrt{N}$, which couple directly to the cavity field $\ac$
through the interaction Hamiltonian $\H_{\mathrm{I}}$. Panel (c) shows
the magnitude of these transverse spin components; in the storage part
it grows as the quantum state is swapped from the cavity and then
decays within $T_2^*$ due to inhomogeneous broadening. Despite the
excitation of very large mean spin components by the $\pi$ pulses, the
much weaker mean values of the stored spin states are recovered as a
primary echo [arrow in panel (c)] and at the final memory
retrieval. Panel (c) also shows the excitation probability $\pexc =
\frac{S_z + N}{2N}$ and the \emph{effective}, $g$-weighted excitation
probability $\pexceff = \frac{1}{2N}(\sum_j
\frac{g_j^2}{\gbar^2}\pauli_z^{(j)}+N)$ versus time. The latter
reaches 89 \% between inversion pulses and levels off at 8 \% after
the second inversion pulse.

The above results can be extracted from mean-value equations alone and
demonstrate the feasibility of the spin ensemble as a classical
memory. In order to assess the quantum properties of the memory we
also calculate the evolution of variances by the coupled first- and
second-moment equations detailed in the Supplementary Material
\cite{SuppMater}, see Fig.~\ref{fig:Varianve_and_fidelity}. Panel (a)
shows the summed variance of $\Xa$ and $\Pa$, which deviates from the
coherent-state value of unity during inversion pulses. At the memory
retrieval the variance also increases when the cavity is tuned to
resonance with $Q = \Qmax$ due to emission from spins left in the
excited state by a non-perfect inversion process (in analogy to
Ref.~\cite{Ruggiero.PhysRevA.79.053851(2009)}), but most importantly
this excess noise of only 11 \% maintains easily the quantum nature of
the memory. Panel (b) shows the summed variance of the spin
components, $\S_x^{\mathrm{eff}}$ and $\S_y^{\mathrm{eff}}$, which
relaxes almost to the coherent-state value at the memory retrieval.

We stress the indispensable role of the spin-frequency inhomogeneity,
which for a resonant cavity in low-$Q$ mode gives rise to the
effective cooperativity parameter $C = \frac{\gens^2}{\kappamax\Gamma}
\approx 0.38$. According to \cite{Julsgaard.PhysRevA.86.063810(2012)}
this ensures (i) that the excess variance of $\Xa$, $\Pa$,
$\S_x^{\mathrm{eff}}$, and $\S_y^{\mathrm{eff}}$ converge to moderate,
finite values during the resonant, inverted period (see,
e.g.,~Fig.~\ref{fig:Varianve_and_fidelity}(a,b) at
$3\:\micro\mathrm{s} \lesssim t \lesssim 4\:\micro\mathrm{s}$), and
(ii) that mean values of the coupled spin-cavity system observables
relax sufficiently fast from possibly imperfect $\pi$ pulses as
exemplified in the inset of Fig.~\ref{fig:MeanValues}(b). For the
off-resonant cavity the first and second moments of the spin
components are damped on the $T_2^*$ time scale as seen in
Fig.~\ref{fig:MeanValues}(c) prior to the primary echo and in
Fig.~\ref{fig:Varianve_and_fidelity}(b) at $t \approx
4\:\micro\mathrm{s}$, respectively. This is essential for the
performance of the protocol; any reminiscence of the inversion pulses
and excess noise in the spin ensemble must vanish both at the time of
the primary echo and of the memory retrieval.
\begin{figure}[t]
  \centering
  \includegraphics{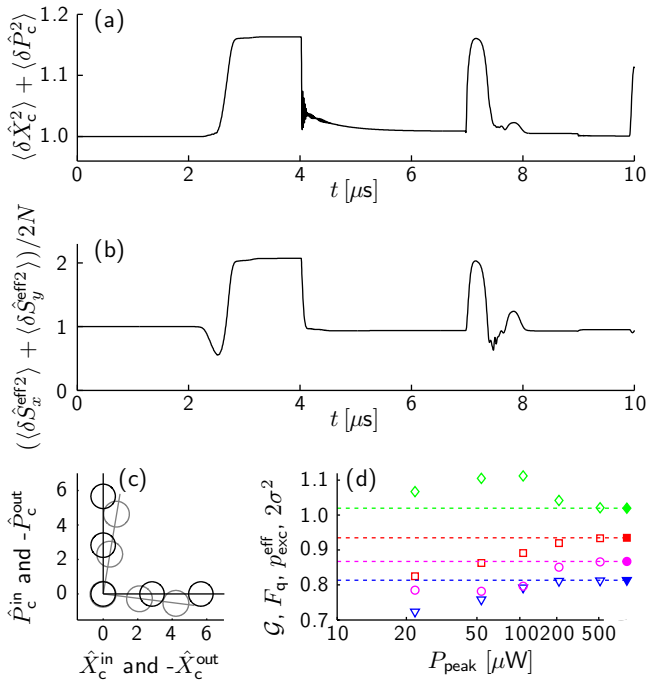}
  \caption{(Color online) (a) Summed variance of cavity-field
    quadratures. (b) The summed, $g$-weighted spin-component variance
    normalized to $2N$. (c) Various input states (black) and output
    states (gray, sign reversed) examined in the protocol. The center
    of circles mark mean values whereas the radii mark the standard
    deviation $\sigma$ of the state. (d) Open symbols: The dependence
    of gain $\mathcal{G}$ (blue triangles), qubit fidelity $\Fq$
    (magenta circles), effective excitation probability $\pexceff$
    (red squares), and summed variance $2\sigma^2$ (green diamonds) on
    the peak power $P_{\mathrm{peak}}$ of the external driving field
    during inversion pulses. Closed symbols: Simulations with
    $P_{\mathrm{peak}} = 100\:\micro$W and homogeneous coupling $g =
    2\pi\cdot 12.5$ Hz, leading to $\mathcal{G} = 0.82$, $2\sigma^2 =
    1.02$, and $\Fq = 87\:\%$.}
\label{fig:Varianve_and_fidelity}
\end{figure}

To assess the performance of the quantum memory, we repeat the above
simulation with various other coherent input states. A selection of
these are shown in Fig.~\ref{fig:Varianve_and_fidelity}(c) in terms of
retrieved mean values and variances (gray circles) as compared to
those of the input states (black circles). We confirm that the
input/output relations constitute a linear map, which (i) essentially
maps vacuum to vacuum (with a slightly increased variance)
demonstrating that the remains of the inversion pulses are negligible
and (ii) which presents a gain factor $\mathcal{G} = 0.79$ for the
mean values. The quadrature variances of the retrieved states amount
to $2\sigma^2 = \mean{\delta\Xa^2} + \mean{\delta\Pa^2} = 1.11$.Since
any quantum state can be expressed as a superposition of coherent
states the memory should work for arbitrary input states,
e.g.~Schr\"odinger cats \cite{Wang.PhysRevLett.103.200404(2009)}, and
qubit states encoded in the $\ket{0}$ and $\ket{1}$ Fock states of the
cavity. The storage time depends on the quantum state and the desired
fidelity. Following \cite{Sherson.Nature.443.557(2006)} we obtain a
qubit fidelity $\Fq = 80$ \% for $\Tmem = 10\:\micro$s.

To investigate the implications of the limited peak power available
for inversion pulses, the above-mentioned analysis is repeated for a
selection of peak powers ranging from 20 {\micro}W to 500 {\micro}W
leading to the results presented in
Fig.~\ref{fig:Varianve_and_fidelity}(d) with open
symbols. Furthermore, a simulation is carried out at
$P_{\mathrm{peak}} = 100\:\micro$W but with a homogeneous distribution
of coupling strengths, $g/2\pi = 12.5$ Hz (solid symbols in
Fig.~\ref{fig:Varianve_and_fidelity}(d)). Clearly, increasing
$P_{\mathrm{peak}}$ presents an increase in $\mathcal{G}$ due to a
better inversion process, but since in an intermediate regime a
fraction of spins experiences a poor inversion process due to
insufficient Rabi frequency (limiting the inversion performance
illustrated by the dashed curve in Fig.~\ref{fig:MeanValues}(c)) we
observe the non-monotonous behavior of $2\sigma^2$ shown in
Fig.~\ref{fig:Varianve_and_fidelity}(d). While increasing driving
powers may be infeasible from an experimental point of view an
alternative route to improvement lies in tailoring a more homogeneous
distribution of coupling strengths, e.g.~by limiting the distance
between NV centers and the cavity.

Continuing the analysis with a homogeneous coupling-strength
distribution (solid symbols in
Fig.~\ref{fig:Varianve_and_fidelity}(d), $\Fq = 87\:\%$), we find the
limiting factors for the obtained fidelity, which in terms of gain can
be written approximately as: $\mathcal{G} =
\mathcal{G}_0\exp(-\kappa[\frac{\pi}{2\gens} +
2T_{\mathrm{chirp}}])\exp(-\gammaperp[\Tmem -
0.7\:{\micro}\mathrm{s}])$. The $\kappa$-dependent factor yields
$\approx 0.92$ due to cavity decay during the resonant swapping
process and during the initial and final frequency chirp of duration
$T_{\mathrm{chirp}}$. The $\gammaperp$-dependent factor yields
$\approx 0.91$ due to spin decoherence (partly suppressed when the
quantum state resides in the cavity or a population degree of
freedom). The main contribution to excess noise arises from imperfect
inversion processes, e.g.~due to the dephasing rate $\gammaperp$
during $\pi$ pulses. In the limit $T_2,\Qmax \rightarrow \infty$ the
qubit fidelity becomes $\approx 97\:\%$, and the origin of the
remaining infidelity ($\mathcal{G}_0 \approx 0.97$ and $2\sigma^2
\approx 1.01$) includes a non-perfect cavity-to-spin transfer (arrow
in Fig.~\ref{fig:MeanValues}(b)) and residual imperfections in the
inversion processes.
\begin{figure}[t]
  \centering
  \includegraphics{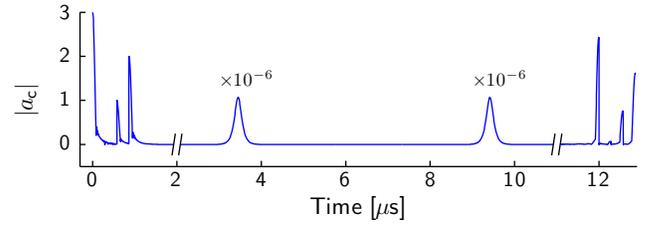}
  \caption{(Color online) The cavity field $|a_{\mathrm{c}}|$ versus
    time in a multi-mode storage example with four input fields
    ($|a_{\mathrm{c}}| = 3, 0, 1, 2$) separated by 0.29 {\micro}s,
    with memory time 12 {\micro}s. The amplitude-cross-talk is below 3
    \%.}
\label{fig:Multimode}
\end{figure}

As demonstrated experimentally for classical pulses
\cite{Wu.PhysRevLett.105.140503(2010)}, the spin-ensemble quantum
memory is multi-mode in nature, which we confirm by simulating storage
and retrieval of four pulses, see Fig.~\ref{fig:Multimode}. The number
of storage modes (proportional to $T_2^*/T_2$) that can be faithfully
addressed and refocused is estimated to be $\sim 100$
\cite{SuppMater}.

In summary, a multi-mode spin-ensemble-based quantum memory for cavity
fields has been proposed and analyzed for a specific realization using
NV centers in diamond. With realistic experimental parameters a qubit
fidelity of $\Fq = 80\:\%$ is predicted for $\Tmem = 10\:\micro$s.
The main limiting processes are clarified, and we predict a qubit
fidelity $\Fq > \frac{2}{3}$, better than achieved by any classical
strategy, for memory times of $\Tmem \lesssim
69\:\micro\mathrm{s}$. The memory time may be further increased by
dynamical decoupling techniques \cite{deLange.Science.330.60(2010),
  Bar-Gill.arXiv.1211.7094} or quantum-state transfer to
nuclear-magnetic degrees of freedom
\cite{Fuchs.NaturePhys.7.789(2011)}.

The authors acknowledge useful discussions with T.~Chaneli\`ere,
D.~Esteve, and Y.~Kubo and support from the EU integrated project
AQUTE, the EU 7th Framework Programme collaborative project iQIT, and
the ANR project QINVC (CHIST-ERA program).

% Uncomment below to make bibliography.
%\bibliography{bibfile}

% Paste .bbl file below:
%

\end{document}

%
%
% Define new commands:
%
\newcommand{\ac}[0]{\ensuremath{\hat{a}_{\mathrm{c}}}}
\newcommand{\adagc}[0]{\ensuremath{\hat{a}^{\dagger}_{\mathrm{c}}}}
\newcommand{\aR}[0]{\ensuremath{\hat{a}_{\mathrm{R}}}}
\newcommand{\betaI}[0]{\ensuremath{\beta_\mathrm{I}}}
\newcommand{\betaR}[0]{\ensuremath{\beta_\mathrm{R}}}
\newcommand{\bra}[1]{\ensuremath{\langle#1|}}
\newcommand{\brag}[0]{\ensuremath{\langle\mathrm{g}|}}
\newcommand{\brae}[0]{\ensuremath{\langle\mathrm{e}|}}
\renewcommand{\c}[0]{\ensuremath{\hat{c}}}
\newcommand{\cdag}[0]{\ensuremath{\hat{c}^{\dagger}}}
\newcommand{\CorrMat}[0]{\ensuremath{\boldsymbol\gamma}}
\newcommand{\Deltacs}[0]{\ensuremath{\Delta_{\mathrm{cs}}}}
\newcommand{\Deltacsmax}[0]{\ensuremath{\Delta_{\mathrm{cs}}^{\mathrm{max}}}}
\newcommand{\Deltacsparked}[0]{\ensuremath{\Delta_{\mathrm{cs}}^{\mathrm{p}}}}
\newcommand{\Deltacstarget}[0]{\ensuremath{\Delta_{\mathrm{cs}}^{\mathrm{t}}}}
\newcommand{\Deltahfs}[0]{\ensuremath{\Delta_{\mathrm{hfs}}}}
\newcommand{\dens}[0]{\ensuremath{\hat{\rho}}}
\newcommand{\epseff}[0]{\ensuremath{\epsilon_{\mathrm{eff}}}}
\newcommand{\erfc}[0]{\ensuremath{\mathrm{erfc}}}
\newcommand{\Fq}[0]{\ensuremath{F_{\mathrm{q}}}}
\newcommand{\gammapar}[0]{\ensuremath{\gamma_{\parallel}}}
\newcommand{\gammaperp}[0]{\ensuremath{\gamma_{\perp}}}
\newcommand{\gbar}[0]{\ensuremath{\bar{g}}}
\newcommand{\gens}[0]{\ensuremath{g_{\mathrm{ens}}}}
\newcommand{\gNV}[0]{\ensuremath{g_{\mathrm{NV}}}}
\renewcommand{\H}[0]{\ensuremath{\hat{H}}}
\renewcommand{\Im}[0]{\ensuremath{\mathrm{Im}}}
\newcommand{\kappac}[0]{\ensuremath{\kappa_{\mathrm{c}}}}
\newcommand{\ket}[1]{\ensuremath{|#1\rangle}}
\newcommand{\ketg}[0]{\ensuremath{|\mathrm{g}\rangle}}
\newcommand{\kete}[0]{\ensuremath{|\mathrm{e}\rangle}}
\newcommand{\lambdag}[0]{\ensuremath{\lambda_{\mathrm{g}}}}
\newcommand{\mat}[1]{\ensuremath{\mathbf{#1}}}
\newcommand{\mean}[1]{\ensuremath{\langle#1\rangle}}
\newcommand{\muB}[0]{\ensuremath{\mu_{\mathrm{B}}}}
\newcommand{\omegac}[0]{\ensuremath{\omega_{\mathrm{c}}}}
\newcommand{\omegas}[0]{\ensuremath{\omega_{\mathrm{s}}}}
\newcommand{\pauli}[0]{\ensuremath{\hat{\sigma}}}
\newcommand{\Pa}[0]{\ensuremath{\hat{P}_{\mathrm{c}}}}
\renewcommand{\Re}[0]{\ensuremath{\mathrm{Re}}}
\renewcommand{\S}[0]{\ensuremath{\hat{S}}}
\newcommand{\Sminuseff}[0]{\ensuremath{\hat{S}_-^{\mathrm{eff}}}}
\newcommand{\Spin}[0]{\ensuremath{\hat{\mathcal{S}}}}
\newcommand{\SpinVec}[0]{\ensuremath{\hat{\boldsymbol{\mathcal{S}}}}}
\newcommand{\Sxeff}[0]{\ensuremath{\hat{S}_x^{\mathrm{eff}}}}
\newcommand{\Syeff}[0]{\ensuremath{\hat{S}_y^{\mathrm{eff}}}}
\newcommand{\tildeac}[0]{\ensuremath{\tilde{a}_{\mathrm{c}}}}
\newcommand{\tildepauli}[0]{\ensuremath{\tilde{\sigma}}}
\newcommand{\Tcaveff}[0]{\ensuremath{T_{\mathrm{cav}}^{\mathrm{eff}}}}
\newcommand{\Techo}[0]{\ensuremath{T_{\mathrm{echo}}}}
\newcommand{\Tmem}[0]{\ensuremath{T_{\mathrm{mem}}}}
\newcommand{\Tres}[0]{\ensuremath{T_{\mathrm{res}}}}
\newcommand{\Tswap}[0]{\ensuremath{T_{\mathrm{swap}}}}
\newcommand{\Tswapres}[0]{\ensuremath{T_{\mathrm{swap}}^{\mathrm{res}}}}
\newcommand{\Var}[0]{\ensuremath{\mathrm{Var}}}
\renewcommand{\vec}[1]{\ensuremath{\mathbf{#1}}}
\newcommand{\Xa}[0]{\ensuremath{\hat{X}_{\mathrm{c}}}}

\section{Supplementary material}
\setcounter{figure}{4} 
% This is what we reached in the main text. The next figure will be Fig. 5.

\subsection{The effective spin-$\frac{1}{2}$ Hamiltonian}
The interaction Hamiltonian between the spin state of NV centers and
the microwave cavity field is calculated in the following. The
electronic part of the Hamiltonian for a single NV-center in an
external magnetic field reads:
\begin{equation}
  \H = D\Spin_z^2 + E(\Spin_x^2 - \Spin_y^2) + \gNV\muB \SpinVec \cdot
    \hat{\vec{B}},
\end{equation}
where $D = 2\pi\cdot 2.88\:\mathrm{GHz}$ is the zero-field splitting
between the $m_{\mathcal{S}}=0$ and $m_{\mathcal{S}} = \pm 1$ states,
$E$ is a splitting between the $m_{\mathcal{S}} = \pm 1$ states
induced by non-axial strain~\cite{Neumann.NewJPhys.11.013017(2009)},
$\gNV = 2$ is the gyro-magnetic ratio, $\muB$ is the Bohr magneton,
$\SpinVec$ is the real spin (not to be confused with the collective
ensemble-spin $\hat{\vec{S}}$ defined elsewhere in the manuscript),
and $\hat{\vec{B}}$ is the external magnetic field. The NV-center axis
defines the direction of quantization along $z$, which we assume in
this work to be parallel to the sample.

The NV center is effectively reduced to a two-level system by applying
a static bias magnetic field $B_z$. If the Zeeman energy verifies
$\gNV\muB B_z \gg E$, which is the case for $B \geq 0.5$\,mT, the spin
states $m_{\mathcal{S}} = 0,\pm 1$ are also energy eigen
states. Taking as our ground state $\ketg = \ket{m_{\mathcal{S}} = 0}$
and excited state $\kete = \ket{m_{\mathcal{S}} = 1}$, the transition
energy becomes: $\omega = D + \gNV\muB B_z +
O\left(\frac{E^2}{\gNV\muB B_z}\right)$. As will be detailed below,
the quantized cavity field is linearly polarized within the $xy$-plane
and can be written: $\hat{\vec{B}} = \delta\vec{B}(\ac + \adagc)$. In
perturbation theory the interaction part of the Hamiltonian then
reads:
\begin{equation}
  \begin{split}
  \H_{\mathrm{I}} &= \gNV\muB \left[\Spin_x\delta B_x(\vec{r})
    + \Spin_y\delta B_y(\vec{r})\right](\ac + \adagc) \\
  &=  \frac{\gNV\muB}{2}\left[
     \Spin_+\{\delta B_x(\vec{r})-i\delta B_y(\vec{r})\} + 
     \Spin_-\{\delta B_x(\vec{r})+i\delta B_y(\vec{r})\}
     \right](\ac + \adagc) \\
  &=  \frac{\gNV\muB}{\sqrt{2}}\left[
     \ac\pauli_+\{\delta B_x(\vec{r})-i\delta B_y(\vec{r})\}+
     \adagc\pauli_-\{\delta B_x(\vec{r})+i\delta B_y(\vec{r})\}
     \right] \\
  &\rightarrow \frac{\gNV\muB|\delta\vec{B}_{\perp}(\vec{r})|}
     {\sqrt{2}}\left[\ac\pauli_+ + \adagc\pauli_-\right].
  \end{split}
\end{equation}
The second line uses the standard definition of raising- and lowering
operators, $\Spin_{\pm} = \Spin_x \pm i\Spin_y$, with properties:
$\Spin_{\pm}\ket{m_{\mathcal{S}}} = \sqrt{\mathcal{S}(\mathcal{S} + 1)
  - m_{\mathcal{S}}(m_{\mathcal{S}}\pm 1)} \ket{m_{\mathcal{S}} \pm
  1}$. Restricted to our effective two-level system, $\ketg$ and
$\kete$, these operators can be stated in terms of Pauli operators:
$\Spin_{\pm} = \sqrt{2}\pauli_{\pm}$, which has been exploited in the
third line together with the rotating-wave approximation. The last
expression is obtained by writing $\delta B_x(\vec{r}) + i\delta
B_y(\vec{r}) = |\delta\vec{B}_{\perp}(\vec{r})| e^{i\theta(\vec{r})}$
and re-defining our basis states as $\ketg$ and
$e^{-i\theta(\vec{r})}\kete$ (such that the phase factor is absorbed
into $\pauli_{\pm}$). We conclude that the coupling between the cavity
field and the effective spin-$\frac{1}{2}$ of the $j$th NV center
located at $\vec{r}_j$ is given by:
\begin{equation}
\label{eq:Calc_gj}
  g_j = \frac{\gNV\muB|\delta\vec{B}_{\perp}(\vec{r}_j)|}{\sqrt{2}}.
\end{equation}
Analytical expressions for the magnetic field of a wave traveling in
the $z$ direction along a coplanar transmission line of the geometry,
shown in Fig.~1(b) of the main text, can be expressed as a function of
the voltage $V_0$ between the center conductor and the ground
plane~\cite{Simons.CoplanarWaveguideCircuits}:
\begin{equation}
  \begin{split}
    B_x &= -\frac{2\mu_0 V_0}{\eta b}\sqrt{\epseff}
    \sum_{n=1}^{\infty}\frac{1}{F_n}\left[\frac{\sin n\pi\delta/2}{n\pi\delta/2}
     \sin\frac{n\pi\bar{\delta}}{2}\right]\cos\frac{n\pi x}{b}
     e^{-\gamma_n y}, \\
    B_y &= -\frac{2\mu_0 V_0}{\eta b}\sqrt{\epseff}
    \sum_{n=1}^{\infty}\left[\frac{\sin n\pi\delta/2}{n\pi\delta/2}
     \sin\frac{n\pi\bar{\delta}}{2}\right]\sin\frac{n\pi x}{b}
     e^{-\gamma_n y}, \\
  \gamma_n &= \sqrt{\left(\frac{n\pi}{b}\right)^2 + \left(\frac{4\pi c b
   \sqrt{\epseff-1}}{n\omegac}\right)^2}.
  \end{split}
\end{equation}
In this expression, $\mu_0$ is the vacuum permeability, $\eta = 376.7$
the vacuum impedance, $c$ the speed of light, $\epseff = (\epsilon_r +
1)/2$, $\epsilon_r$ being the relative dielectric constant of the
substrate, and $\delta = W/b$, $\bar{\delta} = (S+W)/b$, $F_n =
\frac{b\gamma_n}{n\pi}$ are geometrical quantities with $S$ the width
of the coplanar waveguide central conductor, $W$ its distance to each
ground plane [see Fig.~1(b) of the main text], and $b$ the ground
plane width.

In a $\lambda / 2$ coplanar waveguide resonator of length $L$ defined
by a transmission line open at $z=0$ and $z=L$, the fields are
generated by voltages and currents $V(z) = V(0) \cos (\pi z / L)$ and
$I(z) =- i V(0) \sin (\pi z / L) / Z_0$, so that the current in the
middle of the transmission line verifies $I(L/2) = i V(L) / Z_0$, the
same relation as between the current and voltage of a traveling
microwave at a given position of a coplanar waveguide. The zero-point
rms fluctuations of the magnetic field in the middle of the resonator
are thus directly obtained by replacing in the previous expressions
$V_0$ by the zero-point rms fluctuations of the voltage at the
resonator end $\delta V_0 (L) = \omegac \sqrt{\hbar
  Z_0/\pi}$~\cite{Blais.PhysRevA.69.062320(2004)}. In this work we
chose the waveguide geometry so that $Z_0 = 50 \Omega$, as was the
case in a number of recent
experiments~\cite{Kubo.PhysRevLett.105.140502(2010)}.

Using the expression for the rms zero-point magnetic field
fluctuations, we obtain the dependence of the coupling constant of a
spin located in the middle of the resonator ($z=L/2$) as a function of
$x,y$. From that we calculate the coupling constant distribution shown
in Fig.~1(d) of the main text.

\subsection{The electron-spin-resonance transition of the NV ensemble}
We now discuss the spin lineshape and coherence properties that were
chosen in the main text. The electronic spin of all NV centers is also
coupled by hyperfine interaction to the nuclear spin of the nitrogen
atom. This leads to an additional structure of the electron spin
resonance. In the case of NV centers with a $^{14}$N nucleus as is the
case with diamond samples prepared with natural nitrogen abundance,
the spectrum is known to consist of three peaks separated by
$\Deltahfs/2\pi = 2.2$\,MHz~\cite{Felton.PhysRevB.79.075203(2009)}.

In addition to the hyperfine coupling to the nitrogen nucleus, NV
centers are coupled to a bath of other spins that determines their
coherence properties, namely the line width and the spin echo time
$T_2$. Two different well-identified spin baths contribute to NV
center decoherence: nitrogen impurities (so-called P1 centers), and
$^{13}$C nuclear spins. Hybrid circuit experiments usually require a
rather high concentration in NV centers (typically $1$\,ppm or more),
in order to efficiently absorb single photons or equivalently to reach
the strong coupling regime. This high concentration also implies a
comparatively large concentration of P1 centers (on the order of a few
ppm), which at this level are mostly responsible for the sample
coherence properties~\cite{vanWyk.JPhysDApplPhys.30.1790(1997)}. A
sample with $[\mathrm{NV}] \sim 2$\,ppm and $[\mathrm{P1}] \sim
2$\,ppm can be achieved with usual crystal preparation techniques
(electron irradiation of a nitrogen-doped crystal). According
to~\cite{vanWyk.JPhysDApplPhys.30.1790(1997)}, such a sample would
have a line width of each of the hyperfine peaks around $w/2\pi =
0.5$\,MHz, a spin-echo coherence time $T_2 \sim
100\:\micro\mathrm{s}$, and according to
\cite{Kubo.PhysRevLett.107.220501(2011)} would be coupled to a
coplanar resonator with an ensemble coupling constant $\gens/2\pi \sim
3$\,MHz. These are parameters similar to the ones we use in our
calculations (see main text), apart for the line width which we take
somewhat broader $w/2\pi = 2$\,MHz. Broadening the line width
purposely could easily be achieved, either by adding a slightly
inhomogeneous magnetic field, or by a slight misalignment of the
magnetic field with a crystalline axis, which results in a slight
difference in Zeeman shift for NV centers with different orientation
\cite{Kubo.PhysRevLett.107.220501(2011)}. We finally note that in
order to avoid complications of Electron Spin Echo Envelope Modulation
(ESEEM) caused by the interaction with the bath of $^{13}$C
nuclei~\cite{Stanwix.PhysRevB.82.201201R(2010)}, it would be
preferable to use a $^{12}$C-isotope-enriched diamond crystal for this
experiment. With a sample having natural abundance of $^{13}$C, our
refocusing protocol would still work, but only for discrete times
corresponding to the spin-echo revivals depending on the applied
magnetic field, and for NV centers having their axis well aligned with
the magnetic field~\cite{Maze.PhysRevB.78.094303(2008),
  Stanwix.PhysRevB.82.201201R(2010)}. All effects taken together, the
spin distribution can be expressed as
\begin{equation}
  f(\Delta) = \frac{w}{6\pi}\left[\frac{1}{(\Delta-\Deltahfs)^2 + \frac{w^2}{4}}
   + \frac{1}{\Delta^2 + \frac{w^2}{4}} 
   + \frac{1}{(\Delta+\Deltahfs)^2 + \frac{w^2}{4}}\right],
\end{equation}
The corresponding characteristic width
\cite{Julsgaard.PhysRevA.86.063810(2012)} can be derived:
\begin{equation}
  \Gamma =  \left(\gammaperp+\frac{w}{2}\right)
    \frac{(\gammaperp+\frac{w}{2})^2 + \Deltahfs^2}
         {(\gammaperp+\frac{w}{2})^2 + \frac{1}{3}\Deltahfs^2}.
\end{equation}
We note that the free induction decay of an ensemble of identically
prepared spins, $\mean{\pauli_-^{(j)}(0)} \equiv \mean{\pauli_-(0)}$,
can be expressed as:
\begin{equation}
  \begin{split}
  \mean{\S_-(t)} &= \sum_{j=1}^N\mean{\pauli_-^{(j)}(0)}e^{-(\gammaperp + i\Delta_j) t}
   = N\mean{\pauli_-(0)}\int_{-\infty}^{\infty}f(\Delta)
     e^{-(\gammaperp + i\Delta) t}d\Delta \\
  &= \frac{1}{3}\mean{\S_-(0)}[1+2\cos(\Deltahfs t)]e^{-(\gammaperp + \frac{w}{2})t}.
  \end{split}
\end{equation}
As stated in the main text, the time scale for free induction decay,
$T_2^* = \frac{2}{w}$ (taking $\gammaperp \ll w$), is seen to
characterize the envelope function of the modulated decay, caused by
the hyperfine splitting.

\section{Equations of motion for first and second moments using
  sub-ensembles}
With the considerations of the above section, a Hamiltonian for the
physical system can be written in the frame rotating at the central
spin frequency, $\omegas$ as:
\begin{equation}
  \H = \hbar\Deltacs\adagc\ac + \frac{\hbar}{2}\sum_{j=1}^N\Delta_j\pauli_z^{(j)}
      + i\hbar\sqrt{2\kappa}(\beta\adagc - \beta^*\ac)
    + \hbar \sum_{j=1}^N g_j(\pauli_+^{(j)}\ac + \pauli_-^{(j)}\adagc),
\label{eq:Hamiltonian_working}
\end{equation}
where $\Deltacs = \omegac - \omegas$ is the detuning of the
cavity-resonance frequency $\omegac$ from $\omegas$, and $\Delta_j =
\omega_j-\omegas$ with $\omega_j$ being the individual spin-resonance
frequencies. The sums add the contribution from the $N$ spins residing
in the volume considered. An external coherent-state field, $\beta$,
may be used to drive the cavity field $\ac$ through a coupling
capacitor, which gives rise to the cavity-field-decay rate
$\kappa$. The $c$-number $\beta$ is normalized such that $|\beta|^2$
is the incoming number of photons per second.

Decay mechanisms are treated in the Markovian approximation. In our
previous work \cite{Julsgaard.PhysRevA.86.063810(2012)} we employed
the formalism of quantum Langevin equations, which has pedagogical
appeal for understanding the structure of Eq.~(\ref{eq:ddt_CorrMat})
below. However, the present work describes more general states of the
spin ensemble, in which case we prefer the approach of the Lindblad
master equation. This reads: $\frac{\partial\dens}{\partial t} =
\frac{1}{i\hbar}[\H,\dens] + \sum_k\mathcal{D}[\c_k]\dens$ with
$\mathcal{D}[\c_k]\dens = -\frac{1}{2}\cdag_k\c_k\dens
-\frac{1}{2}\dens\cdag_k\c_k + \c_k\dens\cdag_k$, where $\c_1 =
\sqrt{2\kappa}\ac$ accounts for the cavity leakage. In addition, for
the $j$th spin a population decay with rate $\gammapar$ is modeled by
$\c_{2,j} = \sqrt{\gammapar}\pauli_-^{(j)}$ and phase decay with
characteristic time $\tau$ by $\c_{3,j} =
\frac{1}{\sqrt{2\tau}}\pauli_z^{(j)}$. Even though $\gammapar = 0$ in
the main text, it is included here for generality.

Our aim is to describe both the storage of weak cavity fields in the
spin ensemble and the impact of very strong driving fields needed for
spin-refocusing schemes. To this end we develop a formalism, which is
applicable for any saturation level of the spins and which accounts
for the first and second moments of relevant physical operators
[Eq.~(\ref{eq:Relevant_operators})]. The inhomogeneous spin ensemble
is divided into $M$ sub-ensembles, $\mathcal{M}_1, \mathcal{M}_2,
\ldots, \mathcal{M}_M$, which can each be regarded as homogeneous with
coupling strength $g_m$, spin resonance frequency $\Delta_m$, and
containing $N_m$ spins for $m = 1,\ldots,M$.

Next, the dynamical variables are described in terms of the
real-valued operators:
\begin{equation}
\label{eq:Relevant_operators}
  \begin{split}
  \Xa &= \frac{\ac + \adagc}{\sqrt{2}}, \\
  \Pa &= \frac{-i(\ac - \adagc)}{\sqrt{2}}, \\
  \S_x^{(m)} &= \sum_{\mathcal{M}_m}(\pauli_+^{(j)} + \pauli_-^{(j)}), \\
  \S_y^{(m)} &= -i\sum_{\mathcal{M}_m}(\pauli_+^{(j)} - \pauli_-^{(j)}), \\
  \S_z^{(m)} &= \sum_{\mathcal{M}_m}\pauli_z^{(j)}.
  \end{split}
\end{equation}
The $\Xa$ and $\Pa$ operators describe the quadratures of the cavity
field with $[\Xa,\Pa] = i$, while the $\S_k^{(m)}$ components
correspond to twice the total spin in each sub-ensemble with
$[\S_{j}^{(m)},\S_{k}^{(m)}] = 2i\epsilon_{jkl}\S_{l}^{(m)}$. These
operators are now linearized around their mean values: $\Xa =
X_{\mathrm{c}} + \delta\Xa$, $\Pa = P_{\mathrm{c}} + \delta\Pa$, and
$\S_k^{(m)} = S_k^{(m)} + \delta\S_k^{(m)}$ for $k = x,y,z$, i.e.~by
definition, $\mean{\delta\Xa} = 0$, etc. Using the master equation,
the following mean-value equations can then be deduced:
\begin{align}
\label{eq:Mean_val_eq_Xa}
  \frac{\partial X_{\mathrm{c}}}{\partial t} &= -\kappa X_{\mathrm{c}}
    +\Deltacs P_{\mathrm{c}} -\sum_m \frac{g_m}{\sqrt{2}}S_y^{(m)}  
    + 2\sqrt{\kappa}\betaR, \\
\label{eq:Mean_val_eq_Pa}
  \frac{\partial P_{\mathrm{c}}}{\partial t} &= -\kappa P_{\mathrm{c}}
    -\Deltacs X_{\mathrm{c}} -\sum_m\frac{g_m}{\sqrt{2}}S_x^{(m)}  
    +2\sqrt{\kappa}\betaI, \\
\label{eq:Mean_val_eq_Sx_kl}
  \frac{\partial S_x^{(m)}}{\partial t} &= -\gammaperp S_x^{(m)}
    -\Delta_m S_y^{(m)} -\sqrt{2}g_m(S_z^{(m)} P_{\mathrm{c}} 
    + \mean{\delta\S_z^{(m)}\delta\Pa}), \\
   \label{eq:Mean_val_eq_Sy_kl}
  \frac{\partial S_y^{(m)}}{\partial t} &= -\gammaperp S_y^{(m)}
    +\Delta_m S_x^{(m)} 
    -\sqrt{2}g_m(S_z^{(m)} X_{\mathrm{c}} 
      + \mean{\delta\S_z^{(m)}\delta\Xa}), \\
   \label{eq:Mean_val_eq_Sz_kl}
  \frac{\partial S_z^{(m)}}{\partial t} &= \sqrt{2}g_m(S_x^{(m)}P_{\mathrm{c}} 
    +\mean{\delta\S_x^{(m)}\delta\Pa} + S_y^{(m)} X_{\mathrm{c}} 
    + \mean{\delta\S_y^{(m)}\delta\Xa}) \\
  \notag &\quad  - \gammapar(S_z^{(m)} + N_m),
\end{align}
where $\betaR = \Re\{\beta\}$ and $\betaI = \Im\{\beta\}$ and
$\gammaperp = \frac{1}{T_2} = \frac{1}{\tau} +
\frac{\gammapar}{2}$. We note that
Eqs.~(\ref{eq:Mean_val_eq_Sx_kl})-(\ref{eq:Mean_val_eq_Sz_kl}) contain
covariances between the spin and cavity-field operators. In practice,
for all calculations in the main text, these are very small and can be
omitted such that the mean-value equations by themselves form a closed
set.

In total, there are $2+3M$ operators to consider (two quadratures of
the cavity field and three spin components for each sub-ensemble),
which can be written on vector form, $\hat{\vec{y}}$, with $\Xa$ and
$\Pa$ as the first two entries and $\S_x^{(1)}$, $\S_y^{(1)}$,
$\S_z^{(1)}$, $\S_x^{(2)}$, $\S_y^{(2)}$, $\S_z^{(2)}$, etc., on the
following entries. The covariance between any two of these operators,
$\hat{y}_k$ and $\hat{y}_l$, is defined as $C(\hat{y}_k,\hat{y}_l) =
2\Re\{\mean{\delta\hat{y}_k \delta\hat{y}_l}\}$, and in particular
$C(\hat{y}_k,\hat{y}_k) = 2\Var(\hat{y}_k)$. These are grouped into a
$(2+3M)\times(2+3M)$ covariance matrix $\CorrMat$ for which the entry
of the $k$th row and $l$th column is $C(\hat{y}_k,\hat{y}_l)$. The
time-evolution of $\CorrMat$ is:
\begin{equation}
\label{eq:ddt_CorrMat}
  \frac{\partial\CorrMat}{\partial t} = \mat{M}\CorrMat 
     +\CorrMat\mat{M}^{\mathrm{T}} + \mat{N},
\end{equation}
where the driving matrix $\mat{M}$ is given by:
\begin{equation}
  \mat{M} =
  \begin{bmatrix}
    \mat{A} & \mat{B}^{(1)} & \mat{B}^{(2)} & \ldots & \mat{B}^{(M)} \\
    \mat{C}^{(1)} & \mat{D}^{(1)} & 0 & \ldots & 0 \\
    \mat{C}^{(2)} & 0 & \mat{D}^{(2)} & \ldots & 0 \\
    \vdots & \vdots & \vdots & \ddots & \vdots \\
    \mat{C}^{(M)} & 0 & 0 & \ldots & \mat{D}^{(M)}
  \end{bmatrix},
\end{equation}
with
\begin{equation}
  \begin{split}
  \mat{A} &=
  \begin{bmatrix}
    -\kappa &  \Deltacs \\ -\Deltacs & -\kappa
  \end{bmatrix}, \quad
  \mat{B}^{(m)} =
  \begin{bmatrix}
    0 & -\frac{g_m}{\sqrt{2}} & 0 \\ -\frac{g_m}{\sqrt{2}} & 0 & 0
  \end{bmatrix}, \\
  &\mat{C}^{(m)} =
  \begin{bmatrix}
    0 & -\sqrt{2}g_m S_z^{(m)} \\
    -\sqrt{2}g_m S_z^{(m)} & 0 \\
    \sqrt{2}g_m S_y^{(m)} & \sqrt{2}g_m S_x^{(m)}
  \end{bmatrix}, \\
  &\mat{D}^{(m)} =
  \begin{bmatrix}
    -\gammaperp & -\Delta_m & -\sqrt{2}g_m P_{\mathrm{c}} \\
    \Delta_m & -\gammaperp & -\sqrt{2}g_m X_{\mathrm{c}} \\
    \sqrt{2}g_m P_{\mathrm{c}} & \sqrt{2}g_m X_{\mathrm{c}} & -\gammapar
  \end{bmatrix}.
  \end{split}
\end{equation}
The noise matrix $\mat{N}$ is block diagonal:
\begin{equation}
  \begin{split}
  \mat{N} &=
  \begin{bmatrix}
    \mat{V} & 0 & \ldots & 0 \\
    0 & \mat{U}^{(1)} & \ldots & 0 \\
    \vdots & \vdots & \ddots & \vdots \\
    0 & 0 & \ldots & \mat{U}^{(M)}
  \end{bmatrix}, \quad
  \mat{V} =
  \begin{bmatrix}
    2\kappa & 0 \\ 0 & 2\kappa
  \end{bmatrix}, \\
  &\mat{U}^{(m)} =
  \begin{bmatrix}
    4\gammaperp N_m & 0 & 2\gammapar S_x^{(m)} \\
    0 & 4\gammaperp N_m & 2\gammapar S_y^{(m)} \\
    2\gammapar S_x^{(m)} & 2\gammapar S_y^{(m)} & 4\gammapar(S_z^{(m)} + N_m)
  \end{bmatrix}.
  \end{split}
\end{equation}
The above derivation of Eq.~(\ref{eq:ddt_CorrMat}) follows from the
master equation with the approximation that moments of order higher
than two have been neglected. Since second moments have proved
negligible in
Eqs.~(\ref{eq:Mean_val_eq_Sx_kl})-(\ref{eq:Mean_val_eq_Sz_kl}), it can
be expected in succession that moments of order higher than two are
also negligible and the approximation is well justified.

Note that both $\mat{M}$ and $\mat{N}$ contain mean values, and hence
the second-moment equation do \emph{not} form a closed set. However,
if all spins are close to the ground or excited state, $S_z^{(m)}
\approx \pm N_m$, the $\S_z^{(m)}$-operators are effectively removed
from the dynamics and the first- and second-moment equations reduce to
the linear, de-coupled sets published in
\cite{Julsgaard.PhysRevA.86.063810(2012)}.

\section{Implementing the quantum memory protocol}
\begin{figure}[t]
  \centering
  \includegraphics[width=12cm]{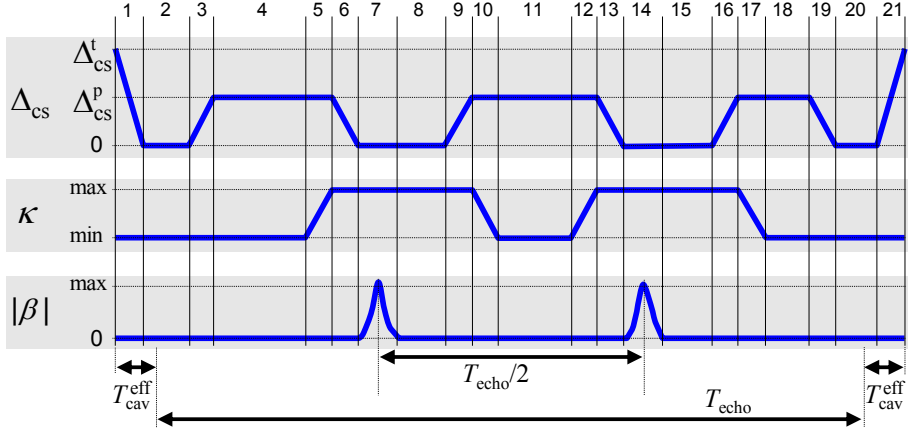}
  \caption{The detailed pulse sequence (not to scale) showing the
    evolution of the external control parameters, $\Deltacs$,
    $\kappa$, and $\beta$. $T_{\mathrm{cav}}^{\mathrm{eff}}$ marks the
    time, which the quantum state spends \emph{effectively} in the
    cavity.}
  \label{fig:PulseSequence}
\end{figure}
In the following the timing details of the quantum memory protocol are
stated. The schematic setup shown in Fig.~2(a) of the main text is
presented in further detail in Fig.~\ref{fig:PulseSequence}. The
protocol sequence consists of 21 parts, and the duration of the $j$th
part is denoted by $T_j$. There are three important cavity-frequency
detunings in the protocol. (1) At the ``target frequency''
$\Deltacstarget = 2\pi\cdot 100$ MHz the cavity field is given to us
for storage or we must deliver the state at this frequency after
memory retrieval, (2) at resonance $\Deltacs = 0$ the spin-cavity
interaction is effective, and (3) at $\Deltacsparked = 2\pi\cdot 50$
MHz the cavity is parked when the spin-cavity interaction needs to be
turned off (in safe distance from the target frequency). Maintaining a
chirp rate of $2\pi\cdot 10$ MHz/ns, we denote by
$T_{\Delta}^{\mathrm{p}} = 5$ ns the time of the chirp $0 \rightarrow
\Deltacsparked$ and by $T_{\Delta}^{\mathrm{t}} = 10$ ns the time of
the chirp $0 \rightarrow \Deltacstarget$ (equal to
$T_{\mathrm{chirp}}$ in the main text). Hence, in the protocol we take
$T_1 = T_{21} = T_{\Delta}^{\mathrm{t}}$ and $T_3 = T_6 = T_9 = T_{13}
= T_{16} = T_{19} = T_{\Delta}^{\mathrm{p}}$. The cavity-field decay
rate $\kappa$ is also varied linearly in time with a duration taken to
be $T_5 = T_{10} = T_{12} = T_{17} \equiv T_{\kappa} = 10$ ns.

The duration $T_2 = T_{20} = \Tswap = 73.7$ ns of the storage and
retrieval is determined numerically by the condition that the cavity
field amplitude $|\mean{\ac}|$ is minimized after part 3 of the
sequence. The ``ideal'' time of $\frac{\pi}{2\gens} = 71.4$ ns is
modified due to cavity decay, inhomogeneous broadening, and the fact
that the information transfer extends slightly into parts 1 and 3 (or
parts 19 and 21).

The duration of the external driving pulses are chosen to be $T_7 =
T_{14} \equiv T_{\pi} = 1\:\micro\mathrm{s}$. The desired
intra-cavity-field mean value $a_{\mathrm{c}}$ is obtained by
tailoring the external driving field $\beta$ as follows: Solve first
Eqs.~(\ref{eq:Mean_val_eq_Sx_kl})-(\ref{eq:Mean_val_eq_Sz_kl}) with
$X_{\mathrm{c}}$ and $P_{\mathrm{c}}$ given by the desired
$a_{\mathrm{c}}$ (the secant hyperbolic pulse stated in the main text)
and with initial conditions $S_x^{(m)} = S_y^{(m)} = 0$ and $S_z^{(m)}
= \pm N_m$ where plus and minus are chosen for the second and first
$\pi$ pulse, respectively. Second, the external $\beta$ can be found
from re-arranging Eqs.~(\ref{eq:Mean_val_eq_Xa})
and~(\ref{eq:Mean_val_eq_Pa}) [we write it on complex form]:
\begin{equation}
  \beta = \frac{1}{\sqrt{2\kappa}}\left[
    \frac{\partial a_{\mathrm{c}}}{\partial t}  
    +(\kappa+i\Deltacs) a_{\mathrm{c}}
    +i\sum_m g_m S_-^{(m)}\right],
\end{equation}
where $S_-^{(m)} = \frac{S_x^{(m)}-iS_y^{(m)}}{2}$ is the solution
from the first step. The first term $\frac{\partial
  a_{\mathrm{c}}}{\partial t}$ accounts for the finite response time
of the cavity while the third term including $S_-^{(m)}$ describes the
reaction field from the magnetic dipole moments of the spin ensemble.

After each inversion pulse the mean values of the cavity field
quadratures, $X_{\mathrm{c}}$ and $P_{\mathrm{c}}$, and of the
transverse spin components, $S_x$ and $S_y$, will inevitably contain
small, non-zero values due to a non-perfect inversion process. These
mean values must relax to appropriate levels before the cavity
parameters can be changed (due to saturation effects in the
cavity-tuning circuitry), which is ensured by (parts 8 and 15) during
the additional time on resonance, $T_8 = T_{15} \equiv \Tres =
1\:\micro\mathrm{s}$. Moreover, these mean values together with the
possible excited-state excess variance must be reduced even further:
If the stored cavity field was the vacuum state, the spin-mode coupled
to the cavity must also be effectively in the vacuum state at the time
of the primary spin echo in part 11 and at the time of the memory
retrieval. Otherwise, any stored quantum state will be
distorted. Specifically, for the second inversion pulse the time
$T_{15}+\ldots+T_{19}$ must be long enough for the mean values to
decay and the time $T_{16}+\ldots+T_{19}$ must be long enough for the
excess noise to decay (a similar condition applies after the first
inversion pulse).

Since relaxation is required after both the first and second inversion
pulses, it is convenient to arrange the sequence such that the primary
spin echo occurs right in the middle of the inversion pulses. This is
obtained when the total time before the first inversion pulse equals
the duration after the second, $T_1+\ldots+T_6 =
T_{15}+\ldots+T_{21}$, which with the choices above reduces to:
\begin{equation}
\label{eq:Relate_T4_and_T18}
    T_4 = \Tres + T_{18}.  
\end{equation}
Next, the duration $\Techo$, between the \emph{effective} focus points
of the reversible spin dephasing process, must fulfill (see bottom
part of Fig.~\ref{fig:PulseSequence}):
\begin{equation}
\label{eq:Techo}
  \Techo = \Tmem - 2\Tcaveff = 2\sum_{j=7}^{13}T_j,
\end{equation}
where the last expression is equal to twice the time between the
inversion pulses. Finally, the total memory time $\Tmem$ is given by
us:
\begin{equation}
\label{eq:Tmem}
  \Tmem = \sum_{j=1}^{21}T_j.
\end{equation}
The above expressions~(\ref{eq:Relate_T4_and_T18})-(\ref{eq:Tmem})
present four equations with the four unknowns $T_4$, $T_{11}$,
$T_{18}$, and $\Techo$. After some algebra we find:
\begin{align}
  T_{11} &= \frac{\Tmem}{2} - 2T_{\Delta}^{\mathrm{p}} - 2T_{\kappa} - T_{\pi}
         - \Tres - \Tcaveff, \\
  T_{18} &= \frac{\Tmem}{4} - T_{\Delta}^{\mathrm{t}} - 2T_{\Delta}^{\mathrm{p}}
         - T_{\kappa} - \frac{T_{\pi}}{2} - \Tres - \Tswap + \frac{\Tcaveff}{2},
\end{align}
and $T_4$ and $\Techo$ follow immediately from
Eqs.~(\ref{eq:Relate_T4_and_T18}) and~(\ref{eq:Techo}). The effective
time $\Tcaveff$ can be calculated by choosing some initial guess,
e.g.~halfway into part 2, and extending $T_{18}$ such that the time
$t_{\mathrm{revival}}$ of the spin revival occurs within part 18. Then
the correct value follows: $\Tcaveff = t_{\mathrm{revival}} - \Techo$.

The multi-mode version of the protocol discussed around Fig.~4 in the
main text has been simulated as follows: After the initial storage
(parts 1-3) and a waiting time of $0.19\:\micro$s, the cavity is
chirped to the target frequency $\Deltacstarget$ and the cavity field
is then \emph{replaced} by a new coherent field ($\mean{\delta\Xa^2} =
\mean{\delta\Pa^2} \rightarrow \frac{1}{2}$, correlations between the
cavity field and each sub-ensemble of spins is set to zero, while
spin-spin correlations are maintained). This new quantum state is then
stored in complete analogy to the first input state. The retrieval is
just the reverse process, and when the cavity reaches the target
frequency $\Deltacstarget$, the cavity is reset to the vacuum
state. This protocol models a perfect, infinitely fast swap between
the cavity and some ``target system'', which holds the unknown quantum
state in the storage part of the protocol and which is prepared in its
ground state at the point of retrieval. In addition, we used a cavity
$Q$-parameter of $10^3$ between these pulses to facilitate a lower
cross talk. In the simulation shown in Fig.~4 of the main text, the
gain factor is 0.80, being lower by the factor
$e^{-\gammaperp(2\:\micro\mathrm{s})}$ due to the extra memory time as
compared to the case of homogeneous coupling strengths in Fig.~3, and
the variance is maintained at $2\sigma^2 = 1.02$. As mentioned in the
summary of the main text, the experimentally feasible case with
inhomogeneous coupling strength and $P_{\mathrm{peak}} =
100\:\micro\mathrm{W}$ presents non-classical storage capabilities
with $\Tmem$ up to 69 {\micro}s, which can be deduced using the fact
that an additional memory time $\Delta T$ will decrease the gain by
the factor $e^{-\Delta T/T_2}$. Using the pulse separation of 0.29
{\micro}s as in Fig.~4 the protocol will accommodate 100 pulses when
$\Tmem = 69\:\micro\mathrm{s}$.
%
%
%
%
%
%

%\bibliography{bibfile}

%